\documentclass{article}
\newcommand{\bea}{\begin{eqnarray}}
\newcommand{\ena}{\end{eqnarray}}
\begin{document}
\topmargin -1cm

\begin{flushright}
MISC-2015-02\\
\today 
\end{flushright}
\begin{center}
{\large \bf 
The model that predicts the maximal 2-3 mixing and  CP violation for neutrinos is revisited }
\vskip 5mm
{\sc Eiichi Takasugi}\footnote{
E-mail:takasugi@cc.kyoto-su.ac.jp\hskip 2mm takasugi@het.phys.sci..osaka-u.ac.jp} \\
{\em Maskawa Institute, Kyoto Sangyo University, Kamigamo, Kyoto, Japan}
 \\
and\\
{\em Department of Physics, Osaka University, Toyonaka, Osaka
 560-0043, Japan}\\
\end{center}

\begin{abstract}
{\em The model of the neutrino mass matrix which we proposed in 2000 is revisited in the light of the recent T2K experiments. This model has the special property that it 
predicts the maximal 2-3 mixing and CP violation under the some simple 
condition.  In this model, if the  condition is relaxed, the 2-3 angle and 
the CP violation deviate from their maximal values and they are related. 
We present such relations for typical cases. }
\end{abstract}



\section{Introduction}
In view of the recent T2K experiments[1],  the mode predicting 
the maximal 2-3 mixing and CP violation attracts  
more attention and seems to require further investigation. In 2000, 
we worked a series of papers[2,3,4] where the neutrino mass 
matrix is proposed in the mass eigenstate basis of charged leptons and 
we made the following two findings. \\
One is that this neutrino mass matrix predicts  
the maximal 2-3 mixing and the maximal CP violation under the some 
simple condition. This is due to the fact that elements of the derived 
neutrino mixing matrix satisfy the special relation which    
is the same as the one discussed by Grimus and Lavoura[6] in 2003\footnote{
I obtained the information about the recent discussions about the neutrino 
mass matrix from the review article by King {\it et al.,}[5].}. 
Also, this neutrino mass matrix has the 
same property of the one discussed by Ma[7], recently. \\
Second is that we found a new mixing 
which is essentially the same as the one which is now called the 
TriBi-maximal mixing proposed by Harrison, Perkins and Scott[8] 
in 2002.These points are explained in the Sec.2. 
In Sec.3, we relax the condition and discuss how the 2-3 mixing and 
the CP violation deviate  from the maximal values and derive their 
relations for some typical cases. In Sec.4, we discuss what neutrino 
mass matrices realize the situation discussed in Sec.3. 
The concluding remarks are given in Sec.5.   

\section{The brief survey of our papers}
In our papers[2,3,4], we proposed the neutrino mass matrix 
in the diagonal basis of the charged lepton mass matrix. \\
\noindent
1) The neutrino mass matrix which we proposed \\
The neutrino mass matrix was given by the combination of 
$S_i$ and $T_i$ as\footnote{In the paper [2], the mass matrix is given by 
$m_\nu=\Sigma_{i}(m_i^0 S_i+\tilde{m_i}T_i)$ which is equivalent to the present expression.}
\begin{eqnarray}
m_\nu=m_1^0 S_1+m_2^0 S_2+m_3^0 S_3+\tilde{m_1}(T_1-S_1)+\tilde{m_2}
(T_2-S_2)+\tilde{m_3}(T_3-S_3),
\end{eqnarray}
where\footnote{These matrices are obtained from $S_3$ and
 $T_3$ by the transformation $PS_i P=S_{i+1}$ and $PT_i P=T_{i+1}$
(mod 3) with $P=T_2$.} 
\begin{eqnarray}
&&
S_1=\frac{1}{3}\pmatrix{1&\omega^2&\omega\cr \omega^2&\omega&1\cr
     \omega&1&\omega^2\cr},\;\;\;\;
S_2=\frac{1}{3}\pmatrix{1&\omega&\omega^2\cr \omega&\omega^2&1\cr
     \omega^2&1&\omega\cr},\;\;\;\;
S_3=\frac{1}{3}\pmatrix{1&1&1\cr 1&1&1\cr 1&1&1\cr},\;\;\;\;
\nonumber\\
&& T_1=\pmatrix{1&0&0\cr 0&\omega&0 \cr 0&0& \omega^2\cr},\;\;\;\;\;
 T_2=\pmatrix{1&0&0\cr 0&\omega^2&0 \cr 0&0& \omega\cr},\;\;\;\;\;
  T_3=\pmatrix{1&0&0\cr 0&1&0 \cr 0&0&1 \cr}.\;\;\;\;\;
\end{eqnarray}
In general, mass parameters $m_i^0$ and $\tilde{m_i}$ are complex and then 
$m_\nu$ becomes a general complex symmetric matrix. \\
Then we found 
that if the mass matrix is transformed by the Tri-maximal mixing matrix 
$V_T$, 
\begin{eqnarray}
V_T=\frac{1}{\sqrt{3}}
\pmatrix{1&1&1\cr \omega&\omega^2&1\cr \omega^2&\omega&1},
\end{eqnarray}
with $\omega=e^{2\pi/3}$, the transformed neutrino mass matrix has the following form,
\begin{eqnarray}
m_\nu'=V_T^T m_\nu V_T=
\pmatrix{m_1^0&\tilde{m_3}&\tilde{m_2}\cr
 \tilde{m_3}&m_2^0&\tilde{m_1}\cr
  \tilde{m_2}&\tilde{m_1}&m_3^0\cr}.
\end{eqnarray}
From this, we found that 
{\bf if all mass parameters, $m_i^0$ and $\tilde{m_i}$ are real}, $m_\nu'$  
becomes  a real symmetric matrix and is diagonalized by a real orthogonal matrix. 
As a result, the neutrino mass matrix $m_\nu$ 
is diagonalized by the matrix which is the Tri-maximal matrix multiplied by 
a real orthogonal matrix $O$, 
i.e., $V=V_TO$. Next, we showed that this matrix has a property
\begin{eqnarray}
V_{2i}=V_{3i}^*\;\;\;\;\;(i=1,2,3).
\end{eqnarray}
and discussed that since by the phase redefinition of the mixing matrix, $V$ can 
converted to the PMNS matrix[9] and the condition in Eq.(5) leads to the constraint 
$|(U_{PMNS})_{2i}|=|(U_{PMNS})_{3i}|$ from which we found[2]      
\begin{eqnarray}
s_{23}^2=c_{23}^2,\;\;\; \cos \delta_{CP}=0,
\end{eqnarray}
where $c_{ij}=\cos \theta_{ij}$ and $s_{ij}=\sin \theta_{ij}$. That is, the maximal 2-3 mixing, 
$\theta_{23}=\pi/4$ and the maximal CP violation phase, $\delta_{CP}=\pm\pi/2$. It is noted 
that the condition in Eq.(5) is exactly the same as the one discussed 
by Grimus and Lavoura[6]. 
\\ 
We note that under our assumption that all mass parameters are real, 
the mass matrix $m_\nu$ becomes a matrix which has the property,
$(m_\nu)_{11}$ and $(m_\nu)_{23}$ are real, and  
$(m_\nu)_{22}=(m_\nu)_{33}^*$, $(m_\nu)_{12}=(m_\nu)_{13}^*$.  
This property of the neutrino mass matrix is exactly the same as the one 
discussed by Ma[7], recently.  
 \\
\vskip 0.2mm
\noindent
2) The TriBi-maximal mixing\\
In our papers[3,4], we found a new mixing by taking a special orthogonal 
rotation for $O$(see Eqs.(28), (33) and (41) in paper[3], 
also the equation in Section 3 in paper[4]) 
\begin{eqnarray}
V_T\pmatrix{1&0&0\cr 0&\frac{1}{\sqrt{2}}&-\frac{1}{\sqrt{2}}\cr
0&\frac{1}{\sqrt{2}}&\frac{1}{\sqrt{2}}}=
\pmatrix{1&0&0\cr 0&\omega &0\cr 0&0&\omega^2\cr}
\pmatrix{\frac{1}{\sqrt{3}}&-\sqrt{\frac{2}{3}}&0\cr
\frac{1}{\sqrt{3}}&\frac{1}{\sqrt{6}}&-\frac{1}{\sqrt{2}}\cr
\frac{1}{\sqrt{3}}&\frac{1}{\sqrt{6}}&\frac{1}{\sqrt{2}}\cr}
\pmatrix{1&0&0\cr 0&-1&0\cr 0&0&i}
\end{eqnarray}
which gave the predictions $\sin^2 2\theta_{sol}=8/9$ and 
$\sin^2 2\theta_{atm}=1$. 
Since neutrino masses are free parameters in this model,  
the mixing matrix is essentially the one called 
TriBi-maximal mixing, i.e., if we interchange  
the 1st and the 2nd columns, we reach to it.\\
We emphasize that if the maximal 2-3 mixing and  
CP violation are confirmed by the experimental observations 
as the T2K experiment implies, 
the model is perfect because  
other mixing angles and neutrino masses 
are obtained in any precision by choosing mass parameters. 
If the maximality for the 2-3 mixing and 
the CP violation is violated, there needs some modification which we 
discuss in the next section.   
 
\section{The relation between the 2-3 mixing angle and the CP violation phase 
in the modified condition}
If some of mass parameters are complex, then the orthogonal matrix $O$ in the 
above of Eq.(5) becomes the unitary matrix and 
then the 2-3 mixing and the CP violation shift from their maximal values. 
In this section, we discuss this effect 
by changing the orthogonal matrix into the unitary one. \\ 
We start from the basis where the Tri-maximal mixing matrix is converted to the 
TriBi-maximal mixing matrix. Then, we consider two typical cases. 
One is that 1) the 2-3 rotation followed by the 
1-2 rotation, and the other is  2) the 1-3 rotation followed by the 1-2 rotation.  
We assume that the 2-3 (or 1-3) rotation is made by an unitary matrix, 
leaving the 1-2 rotation by a real orthogonal matrix. \\
To reach to the TriBi-maximal mixing, we can use the 2-3 rotation given in Eq.(7) followed by the 
exchange of the 1st and the 2nd columns. Here we use the direct way such that
\begin{eqnarray}
\tilde{V}&\equiv &
V_T\pmatrix{\frac{1}{\sqrt{2}}&0&-\frac{1}{\sqrt{2}}\cr
0&1&0\cr \frac{1}{\sqrt{2}}&0&\frac{1}{\sqrt{2}}}
\nonumber\\
&=&\pmatrix{1&0&0\cr 0&\omega^2 &0\cr 0&0&\omega\cr}
\pmatrix{\sqrt{\frac{2}{3}}&\frac{1}{\sqrt{3}}&0\cr
-\frac{1}{\sqrt{6}}&\frac{1}{\sqrt{3}}&-\frac{1}{\sqrt{2}}\cr
-\frac{1}{\sqrt{6}}&\frac{1}{\sqrt{3}}&\frac{1}{\sqrt{2}}\cr}
\pmatrix{1&0&0\cr 0&1&0\cr 0&0&-i}. 
\end{eqnarray} 
\noindent
1) The case of the 2-3 rotation followed by the 1-2 rotation\\
We consider 
\begin{eqnarray}
V'\equiv \tilde{V}\pmatrix{1&0&0\cr 0&c&se^{-i\rho}\cr 0&-se^{i\rho}& c\cr}
\pmatrix{c'&s'&0\cr -s'&c'&0\cr 0&0&1},
\end{eqnarray}
where $c=\cos \theta$, $c'=\cos \theta'$and $s=\sin \theta$, $s'=\sin \theta'$, 
and we take $c>0$ and $c'> 0$ and assume that $s$ and $s'$ are small from the 
observation that TriBi-maximal mixing reproduces the data fairly well. 
We find $V'={\rm diag}(1,\omega^2,\omega)V{\rm diag}(1,1,-i)$, where
\begin{eqnarray}
V=
\pmatrix{\sqrt{\frac{2}{3}}c'-\frac{cs'}{\sqrt{3}}&
\sqrt{\frac{2}{3}}s'+\frac{cc'}{\sqrt{3}}&i\frac{s}{\sqrt{3}}e^{-i\rho}\cr
-\frac{c'}{\sqrt{6}}-\frac{cs'}{\sqrt{3}}+i\frac{ss'}{\sqrt{2}}e^{i\rho}&
-\frac{s'}{\sqrt{6}}+\frac{cc'}{\sqrt{3}}-i\frac{sc'}{\sqrt{2}}e^{i\rho}&
-\frac{c}{\sqrt{2}}+i\frac{s}{\sqrt{3}}e^{-i\rho}\cr
-\frac{c'}{\sqrt{6}}-\frac{cs'}{\sqrt{3}}-i\frac{ss'}{\sqrt{2}}e^{i\rho}&
-\frac{s'}{\sqrt{6}}+\frac{cc'}{\sqrt{3}}+i\frac{sc'}{\sqrt{2}}e^{i\rho}&
\frac{c}{\sqrt{2}}+i\frac{s}{\sqrt{3}}e^{-i\rho}\cr}.
\end{eqnarray}
From this, we find 
\begin{eqnarray}
s_{13}^2&=&\frac{s^2}{3}\nonumber\\
s_{23}^2c_{13}^2&=&\frac{1}{2}-\frac{s^2}{6}-
\sqrt{\frac{2}{3}}sc\sin\rho,
\end{eqnarray}
which lead to
\begin{eqnarray}
\sin \rho=\frac{{\rm sgn}(s)}{2\sqrt{2}}\frac{\cos 2\theta_{23}c_{13}^2}{s_{13}\sqrt{1-3 s_{13}^2}}.
\end{eqnarray}
Next, we compute Jarlskog invariant
\begin{eqnarray}
J_{CP}&=&{\rm Im}[V_{11}V_{22}V_{12}^*V_{21}^*]\nonumber\\
     &=&V_{11}V_{12}\frac{s\cos\rho}{2\sqrt{3}}\nonumber\\
     &=&s_{23}c_{23}s_{12}c_{12}s_{13}c_{13}^2\sin\delta_{CP}
\end{eqnarray}
where the 2nd equality is computed by using the mixing matrix $V$ and 
the 3rd one is done by using the PMNS matrix in the PDG. Since we 
take $c>0$ and $c'> 0$ and assume that $s$ and $s'$ are small , $V_{11}>0$ and $V_{12}>0$, 
so that we find 
\begin{eqnarray}
\cos \rho={\rm sgn}(s)\sin 2\theta_{23}\sin \delta_{CP},
\end{eqnarray} 
where we used $|s|=\sqrt{3}s_{13}$. From Eqs.(12) and (14), we find 
\begin{eqnarray}
|\cos \delta_{CP}||\tan 2\theta_{23}|=
\sqrt{\frac{c_{13}^4}{8s_{13}^2(1-3s_{13}^2)}-1}.
\end{eqnarray}
That is, $|\cos \delta_{CP}|$ is proportional to $|1/\tan 2\theta_{23}|$ and this 
is the relation between the deviations between $\theta_{23}$ and $\delta_{CP}$ from 
their maximal values. 
It is noted that in the limit of $\theta_{23}=\pi/4$, then, $\sin\delta_{CP}=0$ 
is reproduced. By convention, we take $\cos \rho>0$, then for $s>0$,  
$\sin \delta_{CP}>0$ and for $s<0$, $\sin\delta_{CP}<0$. \\
As for $\theta_{12}$, 
\begin{eqnarray}
\tan\theta_{12}=\frac{\sqrt{2}\tan\theta'+\sqrt{1-3s_{13}^2}}
{\sqrt{2}-\sqrt{1-3s_{13}^2}\tan\theta'}.
\end{eqnarray}
In the limit of $s'=\sin\theta'=0$, this relation reduces to the 
well known one
\begin{eqnarray}
s_{12}^2=1-\frac{2}{3}\frac{1}{c^2_{13}}
\end{eqnarray}
which seems to reproduce the data well, so that  the assumption 
that $|s'|$ is small is valid. \\
\noindent
2) The case of the 1-3 rotation followed by the 1-2 rotation. \\
We consider 
\begin{eqnarray}
V'\equiv \tilde{V}\pmatrix{c&0&se^{-i\rho}\cr 0&1&0\cr 
-se^{i\rho}&0& c\cr}
\pmatrix{c'&s'&0\cr -s'&c'&0\cr 0&0&1},
\end{eqnarray}
where we take $c>0$ and $c'> 0$ and assume that $s$ and $s'$ are small. 
We find $V'={\rm diag}(1,\omega^2,\omega)V{\rm diag}(1,1,-i)$, where
\begin{eqnarray}
V=
\pmatrix{\sqrt{\frac{2}{3}}cc'-\frac{s'}{\sqrt{3}}&
\sqrt{\frac{2}{3}}cs'+\frac{c'}{\sqrt{3}}&i\sqrt{\frac{2}{3}}se^{-i\rho}\cr
-\frac{cc'}{\sqrt{6}}-\frac{s'}{\sqrt{3}}-i\frac{sc'}{\sqrt{2}}e^{i\rho}&
-\frac{cs'}{\sqrt{6}}+\frac{c'}{\sqrt{3}}-i\frac{ss'}{\sqrt{2}}e^{i\rho}&
-\frac{c}{\sqrt{2}}-i\frac{s}{\sqrt{6}}e^{-i\rho}\cr
-\frac{cc'}{\sqrt{6}}-\frac{s'}{\sqrt{3}}+i\frac{sc'}{\sqrt{2}}e^{i\rho}&
-\frac{cs'}{\sqrt{6}}+\frac{c'}{\sqrt{3}}+i\frac{ss'}{\sqrt{2}}e^{i\rho}&
\frac{c}{\sqrt{2}}-i\frac{s}{\sqrt{3}}e^{-i\rho}\cr}.
\end{eqnarray}
From this, we find 
\begin{eqnarray}
s_{13}^2&=&\frac{2s^2}{3}.
\end{eqnarray}
Next, from $s_{23}c_{13}=|V_{23}|$, we find
\begin{eqnarray}
\sin \rho=-\frac{{\rm sgn}(s)}{\sqrt{2}}\frac{\cos 2\theta_{23}c_{13}^2}{s_{13}\sqrt{1-\frac{3}{2} s_{13}^2}}.
\end{eqnarray}
Now, we compute Jarlskog invariant and find similarly to the previous case
\begin{eqnarray}
\cos \rho={\rm sgn}(s)\sin 2\theta_{23}\sin \delta_{CP},
\end{eqnarray}
where we used $|s|=\sqrt{3/2}s_{13}$. From Eqs.(21) and (22), we find 
\begin{eqnarray}
|\cos \delta_{CP}||\tan 2\theta_{23}|=
\sqrt{\frac{c_{13}^4}{2s_{13}^2(1-\frac{3}{2}s_{13}^2)}-1}.
\end{eqnarray}
It is noted that in the limit of $\theta_{23}=\pi/4$, then, $\sin\delta_{CP}=0$ 
is derived. By convention, we take $\cos \rho>0$, then for $s>0$,  
$\sin \delta_{CP}>0$ and for $s<0$, $\sin\delta_{CP}<0$. \\
As for $\theta_{12}$, 
\begin{eqnarray}
\tan\theta_{12}=\frac{\sqrt{2-3s_{13}^2}\tan\theta'+1}
{\sqrt{2-3s_{13}^2}-\tan\theta'}.
\end{eqnarray}
In the limit of $s'=\sin\theta'=0$, this relation reduces to 
the well known relation,
\begin{eqnarray}
s_{12}^2=\frac{1}{3c^2_{13}},
\end{eqnarray}
which seems not to reproduce the data. If $\tan \theta_{12}<1/\sqrt{2}$ as the recent 
data[9] imply, then 
$\tan \theta'<-(\sqrt{2}-1)/(\sqrt{2}+1)\simeq -0.17$, but this size is still small enough 
for our argument. 

\section{The mass parameters which realize the rotation given in 
the previous section}
In this section, we discuss what kind of mass parameters produces the 
transformation given in the previous section.  
At first, we note that the mass matrix after the transformation by $\tilde V$ is 
\begin{eqnarray}
\tilde{m_\nu}\equiv \tilde{V}^T m_\nu \tilde{V}
=\pmatrix{\frac{m_1^0+m_3^0+2\tilde{m_2}}{2}&\frac{\tilde{m_1}+\tilde{m_3}}{\sqrt{2}}
&\frac{m_3^0-m_1^0}{2}\cr
\frac{\tilde{m_1}+\tilde{m_3}}{\sqrt{2}}&m_2^0&\frac{\tilde{m_1}-\tilde{m_3}}{\sqrt{2}}\cr
\frac{m_3^0-m_1^0}{2}&\frac{\tilde{m_1}-\tilde{m_3}}{\sqrt{2}}&
\frac{m_1^0+m_3^0-2\tilde{m_2}}{2}}.
\end{eqnarray}
\\
1)We want to realize the 2-3 rotation case given in Eq.(9). 
If we take $m_3^0=m_1^0$ and choose $m_1^0$ and $\tilde{m_2}$ to be real, 
and also 
\begin{eqnarray}
\tilde{m_1}=\frac1{\sqrt{2}}(d+be^{i\alpha}),\;\;\;
\tilde{m_3}=\frac1{\sqrt{2}}(d-be^{i\alpha}),\;\;\;
\end{eqnarray}
where $b$ and $d$ are real, we obtain the mass matrix
\begin{eqnarray}
\tilde{m_\nu}=\pmatrix{a_1&d&0\cr d&a_2& be^{i\alpha}\cr 
 0&be^{i\alpha}&a_3\cr}.
\end{eqnarray}
where $a_1=m_1^0+\tilde{m_2}$, $a_2=m_2^0$ and $a_3=m_1^0-\tilde{m_2}$ and they are real. \\
For the NH case, we perform the see-saw calculation by assuming $a_3$ is large in comparison with 
others. Then, this gives the 2-3 rotation and we obtain $|b/a_3|\simeq s_{13}$ and  $\alpha\simeq -\rho$.  
The effect to the sub-matrix of the 1st and the 2nd generation 
is the change of  the 2-2 element $a_2$ to $a_2-b^2 e^{2i\alpha}/a_3$. Therefore, the effect to the complex phase  
is the order of $|(b^2/a_2 a_3)\sin 2\alpha|\le 
0.1|\sin 2\alpha|$, because $(b^2/a_2a_3)\le s_{13}^2\sqrt{\Delta m^2_{atm}/\Delta m^2_{sol}}$. 
If the shift from 
$\theta_{23}$ from $\pi/4$ is small, then $\alpha\simeq -\rho\sim 0$, so 
that this effect is small. Therefore, the 1-2 rotation can be considered to be a real 
orthogonal one in a good approximation.  \\
For the IH case, since $a_1\simeq \pm a_2$ are large, so that the effect after the see-saw 
calculation is only to the 3-3 element $a_3$. 
Therefore the real 1-2 rotation is achieved by the above mass matrix. \\
Thus, the matrix given in Eq.(28) gives the rotation in Eq.(9) in a good approximation, so that 
the relation between $|\cos \delta_{CP}|$ and $|\tan 2\theta_{23}|$ is realized 
in a good approximation . \\
2) We want to realize the 1-3 rotation case given in Eq.(18). If we take $\tilde{m_1}=\tilde{m_3}=d/\sqrt{2}$, 
$\tilde{m_2}=(a_1-a_2)/2$ and 
\begin{eqnarray}
m_3^0=\frac{a_1+a_2}{2}+be^{i\alpha},\;\;\; 
m_1^0=\frac{a_1+a_2}{2}-be^{i\alpha},
\end{eqnarray}
where $a_i$ and $b$ are real, then we obtain
\begin{eqnarray}
\tilde{m_\nu}=\pmatrix{a_1&d&be^{i\alpha}\cr d&a_2& 0\cr 
 be^{i\alpha}&0&a_3\cr}.
\end{eqnarray}
The similar argument holds for this case. For the NH case, 
the effect after the see-saw calculation, $a_1$ is shifted to $a_1-(b^2/a_3)e^{i2\alpha}$. 
However, for the 1-2 rotation, the complex phase which contributes to the Dirac CP 
violation phase enters as 
$a_1+a_2-b^2 e^{2i\alpha}/a_3$, so that the same argument as discussed in the previous case 
holds. Also for the IH case, the same argument holds as the previous case. Therefore, 
the above mass matrix will give the preferable rotation given in Eq.(18) in a  good approximation. 

\section{The concluding remarks}
In this note, we revisited our old papers in 2000 and presented the findings given there, that is, 
our mass matrix predicts the maximal 2-3 mixing angle and the maximal Dirac CP violation under 
the condition that all mass parameters are real. If this condition is relaxed, both the 2-3 mixing and 
the CP violation deviate from their maximal ones. We considered two typical cases and obtained the relation 
between these deviations, which are expressed by  
\begin{eqnarray}
|\cos \delta_{CP}||\tan 2\theta_{23}|=k(s_{13})
\end{eqnarray}
where $k$ is the function of $\theta_{13}$ given in Eqs.(15) and (23) for two cases. 
If we take $s_{13}=0.15$[9],  $k=2.1$ for the 2-3 rotation case and $k=4.5$ for the 1-3 rotation case.   
For the case of the small deviation case,  
$\Delta_{CP}\simeq 2k\Delta_{23}$, where $\Delta_{CP}=|\delta_{CP}\pm \pi/2|$ and 
$\Delta_{23}=|\theta_{23}-\pi/4|$. \\ 
Our predictions in Eq.(31)  gives a good test of the model 
if the precision measurement of $\delta_{CP}$ and $\theta_{23}$ are made. \\
For the model building point of view, we explored in what situation these cases are 
realized. In Sec.4, we gave some simple choices of mass parameters which reproduce 
the rotations used in Sec.3 in a good approximation.\\
Finally, we comment that relations in Eqs.(15) and (23) are valid for some other class of models. 
Suppose that we construct the model of neutrino mass matrix which realizes the 
TriBi-maximal mixing without phase matrices appeared in Eq.(8) and then rotate 
the mixing matrix by the unitary matrix discussed 
in the text by adding some small mass terms. This case is realized by the change 
$\rho\to \sigma -\pi/2$. Since the relations are independent of this phase, the relations 
hold for these cases. The no 1-2 rotation cases are discussed by Shimizu and Tanimoto[10]. 
It may be interesting to examine the relation between the CP violation phase and mixing angles 
for general unitary matrix case numerically.

\vskip 4mm
\noindent
{\huge Acknowledgment}\\
I would like to thank Dr. Atsushi Watanabe for giving me the recent information 
on the neutrino physics and also I enjoyed discussions with him.

\end{document}